\documentclass[showpacs,amsmath,nofootinbib,amssymb,noeprint, reprint,prl]{revtex4-1}
\pdfoutput=1
\usepackage{array,multirow}
\usepackage{color}
\usepackage{amsfonts}
\usepackage{bm}

\usepackage{epsfig}
\def\cn{{\mathcal N}}

\def\la{\lambda}
\def\ep{\epsilon}

\def\lb{\bar{\lambda}}

\def\Or[#1]{{\text{O}}\left({#1}\right)}
\def\dotl[#1,#2]{\left\langle #1, #2 \right\rangle}
\def\dotlb[#1,#2]{[ #1, #2 ]}
\def\dotp[#1,#2]{(#1) \cdot (#2)}
\def\n4sym{{\cal N}=4 SYM}
\def\>{\rangle}
\def\<{\langle}
\newcommand{\vect}[1]{{\boldsymbol{#1}}}
\def\ads[#1]{$\text{AdS}_{#1}$}
\usepackage{natbib}
\usepackage{hyperref}
\begin{document}
\title{BCFW for Witten Diagrams}
\author{Suvrat Raju}
\affiliation{Harish-Chandra Research Institute, Chatnag Marg, Jhunsi, Allahabad 211019, India.}
\begin{abstract}
We show that a generalization of the BCFW recursion relations gives a new and efficient
method of computing correlation functions of the stress tensor or conserved currents in conformal
field theories with an \ads[d+1] dual, for $d \geq 4$, in the limit where 
the bulk theory is approximated by tree-level Yang-Mills or gravity. 
In  supersymmetric theories,  additional correlators of operators that live in the same multiplet as a conserved current or stress tensor  can be computed by these means.
\end{abstract}
\preprint{HRI/ST/1009}
\maketitle
\paragraph{\bf Introduction:}
The past few years have seen an intense study of the Britto-Cachazo-Feng-Witten (BCFW) recursion relations for gauge and gravity theories \cite{Britto:2004ap,*Britto:2005fq}. These
recursion relations not only provide an extremely efficient route to S-matrix elements, they also do so without making any explicit reference to a local Lagrangian. Consequently, they are of interest both for computational reasons (such as the computation of S-matrix elements for physics at the Large Hadron Collider) and because they might help us shed new light on the formalism of quantum field theory. 

Hitherto, it has been believed that these techniques apply only to quantum field theories in flat space. In this paper, we point out that the BCFW recursion relations can be generalized to compute sums of tree-level Witten diagrams in quantum field theories in anti-de Sitter (AdS) space. By the celebrated AdS/CFT  correspondence \cite{Maldacena:1997re,*Witten:1998qj,*Gubser:1998bc}, this gives us new recursion relations for correlation functions in the dual conformal field theory (CFT).

The calculation of Witten diagrams  involving bulk gravity  is notoriously difficult
because of the infinite number of interaction vertices that proliferate as the number of external legs increase. As a result, even the four-graviton amplitude
in AdS$_5$ (dual to the four-point correlator of the stress tensor in ${\cn=4}$ Super-Yang-Mills (SYM) at strong coupling) has never been computed directly.
The same problem exists in flat space where 4D gravity has 2850 four-point vertices; the BCFW recursion relations reduce all computations 
down to the calculation of 
the three-point function  and greatly reduce this complexity. Our new recursion relations do the same for gravity in AdS,
and we anticipate that they will simplify the computation of boundary correlators.

The physical intuition underlying this study is as follows. The BCFW recursion relations are predicated on the behaviour of Yang-Mills (YM) and gravity amplitudes when two of the external momenta are stretched off the infinity in a ``complex direction.'' Although this is not strictly 
a high energy limit, it is nonetheless true that the amplitude is dominated by interactions between a soft background and a highly boosted particle at a {\em single point.} In this limit, 
we do not expect this highly boosted particle to see the curvature of the neighboring spacetime region. On the other hand, we do need to integrate over the different points where this interaction can occur. (This is similar to the intuition used in \cite{Polchinski:2001tt}.) This process leads to the modified recursion relations that we present below. A higher-point correlator is broken down into the integral
of the product of two lower-point correlators. Just as in flat space, we can continue this process till we are left only with three-point functions.

If we 
set out to compute 
a vacuum-correlator in the boundary theory, with all normalizable modes switched off in the bulk, the recursion relations
lead us to correlators computed in the presence of specific states; in the bulk this corresponds to turning on some normalizable modes. We will
call these generalized correlators, transition amplitudes. 

A further extension of our recursion relations allows us to compute transition
amplitudes in supersymmetric theories, including ${\cn=4}$ SYM and 
the theory on multiple M5 branes in the supergravity limit. Perturbative computations in supersymmetric theories are often tedious; the recursion relations 
that we present ameliorate this by using
a generalization of Nair's on-shell superspace \cite{Nair:1988bq}.


\paragraph{\bf Review of Perturbation Theory in AdS:}
We will work
in Poincare coordinates where the metric is
\begin{equation}
d s^2 = g_{\mu \nu} d x^{\mu} d x^{\nu} = z^{-2} \left(d z^2 + \eta_{i j} d x^i d x^j \right).
\end{equation}
Poincare invariance in $d$ dimensions makes it convenient to Fourier
transform functions of $x^i$ and we will call the conjugate variables --- $k_i$ --- ``momenta.'' 

For a non-interacting massless scalar field in AdS with spacelike momentum, $\vect{k}$, we have the unique solution, $\phi = \phi_0 e^{i \vect{k} \cdot \vect{x}} z^{\nu} K_{\nu} (|\vect{k}| z),$ where $|\vect{k}| = |\vect{k}^2|^{1/2}$ and $\nu = d/2$, while for timelike momentum, we have the non-normalizable solution $\phi = \phi_0 e^{i \vect{k} \cdot \vect{x}} z^{\nu} Y_{\nu} (|\vect{k}| z),$ and the normalizable solution $\phi = \phi_0 e^{i \vect{k} \cdot \vect{x}} z^{\nu} J_{\nu} (|\vect{k}| z).$  We will use a uniform notation to write these solutions as
$\phi = \phi_0 e^{i \vect{k} \cdot \vect{x}} E_{\nu} (|\vect{k}| z),$ 
where $z^{-\nu} E_{\nu} $ is one of $K_{\nu}, J_{\nu}, Y_{\nu}$. 

The free Yang-Mills equations in AdS are solved by
\begin{equation}
\label{normalsolym}
A_{i}^{\rm a}(\vect{x},z) = \ep_{i}^{\rm a} E_{\nu_1}(|\vect{k}| z) e^{i \vect{k} \cdot \vect{x}}; A_0^{\rm a} = 0, \vect{k} \cdot \vect{\ep}^{\rm a} = 0,
\end{equation}
where $0$ refers to the z-direction, $\nu_1 \equiv \nu - 1$, and the color index, ${\rm a}$, is not italicized.
We can move away from this gauge through $A_{\mu}^{\rm a}(\vect{x},z) \rightarrow A_{\mu}^{\rm a}(\vect{x},z) + \partial_{\mu} \phi^{\rm a},$ where $\phi^{\rm a}$ is any scalar field.
Similarly, freely propagating gravity waves comprise transverse traceless tensors:
\begin{equation}
\label{normalsolgrav}
h_{i j} =  \ep_{i j} z^{-2} E_{\nu}(|\vect{k}| z) e^{i \vect{k} \cdot \vect{x}}; h_{0\mu} = 0, k_i \ep^{i j} = 0, \ep^{i}_{i} = 0.
\end{equation}
We will refer to $\ep_i^{\rm a}$ and $\ep_{i j}$ as ``polarization vectors.'' 

The propagator for scalars, and for gauge-bosons and gravitons in this axial gauge
is ($\int_{k,p} \equiv \int {-i d^d \vect{k} d p^2 \over 2 (2 \pi)^d }$)
\begin{equation}
\label{axialpropagator}
\begin{split}
G^{\text{scal}} = &\int_{k,p} {
e^{i \vect{k} \cdot (\vect{x} - \vect{x'})} 
z^{\nu} J_{\nu}(p z) J_{\nu} (p z') (z')^{\nu} 
\over 
\left(\vect{k}^2 + p^2 - i \epsilon \right)}, \\
G^{\text{YM}}_{i j} = 
&\int_{k,p}  
{e^{i \vect{k} \cdot (\vect{x} - \vect{x'})}  
 (z z')^{\nu_1} J_{\nu_1}(p z) J_{\nu_1} (p z') {\cal T}_{i j} 
\over 
\left(\vect{k}^2 + p^2 - i \epsilon \right)} , \\
G^{{\text{grav}}}_{i j, k l} =
&\int_{k,p} \left[{
e^{i \vect{k} \cdot (\vect{x} - \vect{x'})} 
z^{\nu-2} J_{\nu}(p z) J_{\nu} (p z') (z')^{\nu - 2} \over  
\left(\vect{k}^2 + p^2 - i \epsilon\right)} \right.   \\ &\times {1 \over 2} \left.\left({\cal T}_{i k} {\cal T}_{j l} + {\cal T}_{i l} {\cal T}_{j k} - 
{2 {\cal T}_{i j} {\cal T}_{k l}\over d-1} \right)\right],  
\end{split}
\end{equation}
where ${\cal T}_{i j} = \eta_{i j} + k_{i} k_{j}/p^2$ 
and we have suppressed the trivial color dependence in $G^{\text{YM}}$ \cite{Liu:1998ty}. What will be important for us is that, 
in each case,  at $p^2 = -\vect{k}^2$, the numerator of the integrand breaks up into a sum of a product of normalizable modes.

\paragraph{\bf Transition Amplitudes in AdS:}
Consider CFT operators $O(\vect{k_{3 1}}), \ldots O(\vect{k_{3 n_3}})$ and states $s, s'$ that are dual, respectively, to linear combinations 
of normalizable modes with momenta $\vect{k_{1 1}}, \ldots \vect{k_{1 n_1}}$ and $\vect{k_{2 1}}, \ldots \vect{k_{2 n_2}}$ in
the bulk. We will examine the transition
amplitude
\begin{equation}
\label{transdef}
T(\vect{k_{l m}}) (2 \pi)^{d} \delta^{d}(\sum_{l m} \vect{k_{l m}}) = \langle s | O(\vect{k_{3 1}}) \ldots O(\vect{k_{3 n_3}}) | s' \rangle.
\end{equation}
Physically, we may 
think of $|s'\rangle, \langle s |$ as specifying data along the past and future horizons of the Poincare patch; we are then asking for the probability that
the operators $O(\vect{k_{3 m}})$ will induce a transition between these states. 

We will work at tree-level in bulk perturbation theory. To compute
transition amplitudes, 
we
draw bulk-bulk diagrams as usual. Then, we contract the legs with momenta in the set $\vect{k_{3 m}}$ with
bulk to boundary propagators (non-normalizable modes), and the other legs, which carry  momenta in the set $\vect{k_{1 m}}$ or $\vect{k_{2 m}}$, with normalizable modes. The reader may prefer to think only in terms of this perturbative 
prescription and should consult \cite{Balasubramanian:1999ri,*Balasubramanian:1998de,*Balasubramanian:1998sn} for further discussion. A vacuum correlator is just a special case of a transition amplitude, where all normalizable modes are switched off.

The structure of perturbation theory tells us that transition amplitudes
are produced by the action of a 
multi-linear operator on a set of (normalizable or non-normalizable) solutions to the equations of motion.
For example, in Yang-Mills, with $A^{\rm a_m}_{\mu_m}(\vect{x},z)$ drawn from \eqref{normalsolym}
\begin{equation}
T = G(A^{\rm a_1}_{\mu_1}(\vect{x},z), \ldots A^{\rm a_n}_{\mu_n}(\vect{x},z)).
\end{equation}
These operators obey Ward identities:
\begin{equation}
\label{wardidentity}
G(\nabla_{\mu_1} \phi^{\rm a_1}(\vect{x},z), A^{\rm a_2}_{\mu_2}(\vect{x},z), \ldots A^{\rm a_n}_{\mu_n}(\vect{x},z)) = 0,
\end{equation}
for any  $\phi^{\rm a_1}(\vect{x},z).$ An analogous identity holds for gravity.

Below, we will consider transition amplitudes, $T(\vect{k_m}, \vect{\ep}^{\rm a_m})$, that depend on a set of discrete momenta, but also on polarization vectors for gauge-bosons and gravitons. The reader should note that some of the $\vect{k_m},\vect{\ep}^{\rm a_m}$ may correspond to normalizable modes, and others to non-normalizable modes; this will be left implicit.

\paragraph{\bf BCFW for Scalars:}
\label{secbcfwrecursion}
We start with a  massless scalar $\phi^3$ theory to explain the main idea in a simple setting.
A four-point transition amplitude involves three  terms. 
\begin{equation}
\label{typical}
\begin{split}
&T(\vect{k_1}, \vect{k_2}, \vect{k_3}, \vect{k_4}) =
\int \Bigl[{E_{\nu}(|\vect{k_1}| z_1) E_{\nu}(|\vect{k_2}| z_1)  z_1^{\nu} J_{\nu}(p z_1) \over (\vect{k_1} + \vect{k_2})^2 + p^2}  \Bigr. \\ &\times \Bigl. z_2^{\nu} J_{\nu}(p z_2) E_{\nu}(|\vect{k_3}| z_2) E_{\nu}(|\vect{k_4}| z_2) \Bigr] {i d z_1 d z_2 d p^2 \over 2 (z_1 z_2)^{d+1}} + \ldots,
\end{split}
\end{equation} 
where the $\ldots$ are the $t$ and $u$ channel terms. Now, consider the extension $ \vect{k_1} \rightarrow \vect{k_1} + \vect{q} w, \vect{k_4} \rightarrow \vect{k_4} - \vect{q} w$, which depends on the
parameter $w$ and where 
$\vect{q}^2= \vect{q} \cdot \vect{k_1} = \vect{q} \cdot \vect{k_4} = 0.$
This generically requires $\vect{q}$ to have complex components.
Under this extension, the {\em integrand} of \eqref{typical} is a rational function of $w$ (although, of course, the integral itself is not)
and has a pole at $(2 \vect{q} \cdot \vect{k_2}) w = -(p^2 + (\vect{k_1} + \vect{k_2})^2) $ with
a residue
\begin{equation}
\label{residue}
\begin{split}
{-i\over  4 \vect{q} \cdot \vect{k_2}} &\left[-i E_{\nu}(|\vect{k_1}| z_1) E_{\nu}(|\vect{k_2}| z_1) z_1^{\nu} J_{\nu}(p z_1) z_1^{-d-1}\right] \\
\times &\left[ -i z_2^{\nu} J_{\nu}(p z_2) E_{\nu}(|\vect{k_3}| z_2) E_{\nu}(|\vect{k_4}| z_2) z_2^{-d-1} \right] .
\end{split}
\end{equation}
Each bracketed terms is the integrand for a 3-point function! There is
also a pole at $w = \infty$ in \eqref{typical} because the integrand of the diagram with a 
contact interaction between $\vect{k_1}$ and $\vect{k_4}$ goes to a constant at large $w$. 

It is easy to see that the same structure persists for $n$-point amplitudes. Poles in the integrand of a transition amplitude occur when the denominator of a propagator vanishes; the residue is the product of 
the integrands of two lower-point amplitudes and a simple factor from the propagator. The inclusion of the residue from $w = \infty$ permits us to completely reconstruct the integrand. So,
\begin{equation}
\label{scalarecurs}
\begin{split}
&T(\vect{k_1}, \ldots \vect{k_n}) = {\cal B} + \sum_{\{\pi\},m}\int  {-i {\cal T}^2 \over 2(p^2 + \vect{K}^2)}  d p^2, \\
&{\cal T}^2 \equiv  {T(\vect{k_1}(p), \ldots \vect{k_m'}) T(-\vect{k_m'},\ldots \vect{k_n}(p) ) }.
\end{split}
\end{equation}
 The sum is over all ways of partitioning
the momenta into two sets $\{\vect{k_1},\vect{k_{\pi_2}}, \ldots \vect{k_{\pi_m}}\}, \{\vect{k_{\pi_{m+1}}}, \ldots \vect{k_{n}} \}$, with $\vect{k_1}$ in one and $\vect{k_n}$ in the other. Also, $\vect{K} = \vect{k_1} + \sum_{2}^m \vect{k_{\pi_m}};  w(p)=-(\vect{K}^2 + p^2)/(2 \vect{K} \cdot \vect{q});  \vect{k_1}(p)=\vect{k_1}+ \vect{q} w(p); \vect{k_n}(p) = \vect{k_n} - \vect{q} w(p); \vect{k_m'} = -\vect{K} - \vect{q} w(p)$. The ``boundary term,''  ${\cal B}$, is the contribution from the pole at $w = \infty$, comprising the sum of all diagrams where $\vect{k_1}$ and $\vect{k_n}$ meet at a point.
 
Note that if we set out to compute a vacuum correlator, all the $E$ in \eqref{typical} 
are non-normalizable. Nevertheless, as we see from \eqref{residue}, the mode 
corresponding to $\vect{k_m'}$ in \eqref{scalarecurs} will always be normalizable. This is implicit in \eqref{scalarecurs}.

\paragraph{\bf Yang-Mills:}
Following \cite{ArkaniHamed:2008yf}, we expand the gauge field as ${\cal A}_{\mu}^{\rm a} = A_{\mu}^{\rm a} + a_{\mu}^{\rm a}$. In background field gauge, the quadratic Lagrangian
for $a_{\mu}^{\rm a}$ is
\begin{equation}
\label{gaugeffectact}
2 {\cal L} = D_{\mu} a_{\nu}^{\rm a} D^{\mu} a^{\nu,{\rm a}} + \left(2 F^{\mu \nu, {\rm{a}}}f^{\rm a b c} +  R^{\mu \nu} \delta^{\rm b c}\right) a_{\mu}^{\rm b} a_{\nu}^{\rm c}.
\end{equation}
where $D_{\mu}$ is covariant in spacetime and with respect to the background field;  $F$ is the background field strength; $f$ gives the structure constants and $R$ is the
Ricci tensor.

We examine the large $w$ behaviour of the two-point function for $a_{\mu}^{\rm a}$, 
BCFW extended as above. 
With q-lightcone gauge for the background field, $\vect{q} \cdot  \vect{A}^{\rm a} = 0$, all $\Or[w]$ interactions come from the diagrams where
there is a {\em single interaction} of the fluctuating field with the background. This is because every propagator comes with a factor of $w^{-1}$. (With the use of \eqref{wardidentity}, we can show that the $k_i k_j$ terms in \eqref{axialpropagator} do not spoil this power counting.)

From the effective action \eqref{gaugeffectact}, we see that the dominant contribution to the transition amplitude comes from
\begin{equation}
\label{twopointbcfw}
\begin{split}
\int &\left[A^{\mu,{\rm a}} f^{\rm a b c} \left(a_{1}^{\nu,{\rm b}} \nabla_{\mu} a_{n,\nu}^{\rm c} - a_{n}^{\nu,{\rm c}} \nabla_{\mu} a_{1 \nu}^{\rm b} \right) \right. \\
&\left.
+ 2 F^{\mu \nu,{\rm a}} a_{1 \mu}^{\rm b} a_{n \nu}^{\rm c} f^{\rm a b c} + \Or[{1 \over w}] \right]  {d^{d} \vect{x} dz \over z^{d+1}}, 
\end{split}
\end{equation}
where $\vect{a_1}, \vect{a_n}$ belong to \eqref{normalsolym}.
Below,
we will suppress the color-factors, which are unimportant for our purposes.

We choose the polarization for $\vect{a_1}$ by $\vect{\ep_1} = \vect{q}$, and define $\vect{t}$ by
$ a_{1 \mu} \equiv w^{-1} \left(\partial_{\mu} \phi - t_{\mu}\right),$
where $\phi = e^{i (\vect{k_1} + \vect{q} \omega) \cdot \vect{x}} E_{\nu_1} (z).$
By the Ward identity now, instead of $a_{1 \mu}$, we can use $w^{-1} t_{\mu}$ in \eqref{twopointbcfw}. 
As a result, the terms in the integrand of \eqref{twopointbcfw} die off at large $w$ if (a) $\vect{\ep_n}$ does not grow at large $w$ (which requires $\vect{\ep_n} \cdot \vect{q} = 0$) and (b) $\vect{k_1} \cdot \vect{\epsilon_n} = 0$. In $d=4$ this forces us to take $\vect{\ep_n} = \vect{q}$ also. For $d > 4$, we can choose an $\vect{\ep_n} \neq \vect{q}$ that is orthogonal to $\vect{k_1},\vect{k_n},\vect{q}$.

With this choice of $\vect{\ep_1}=\vect{q}$ and these constraints on $\vect{\ep_n}$,  we can reconstruct the integrand, up to terms that integrate to zero, 
using its poles at finite $w$. Repeating the argument above, we
get the recursion relation
\begin{equation}
\label{gaugerecurs}
\begin{split}
&T(\vect{k_1},\vect{\ep_1}, \ldots \vect{k_n},\vect{\ep_n}) = \sum_{\{\pi\},m,\vect{\ep_m'}} \int   {-i {\cal T}^2 \over 2 (p^2 + \vect{K}^2)} d p^2,\\
&{\cal T}^2 \equiv {T(\vect{k_1}(p),\vect{\ep_1}, \ldots \vect{k_m'},\vect{\ep_m'}) T(-\vect{k_m'},\vect{\ep_m'}, \ldots \vect{k_n}(p),\vect{\ep_n} )}.
\end{split}
\end{equation}
This has no boundary term and the sum now also runs over all normalized polarization vectors for $\vect{k_m'}$.

\begin{figure}
\includegraphics[height=2cm]{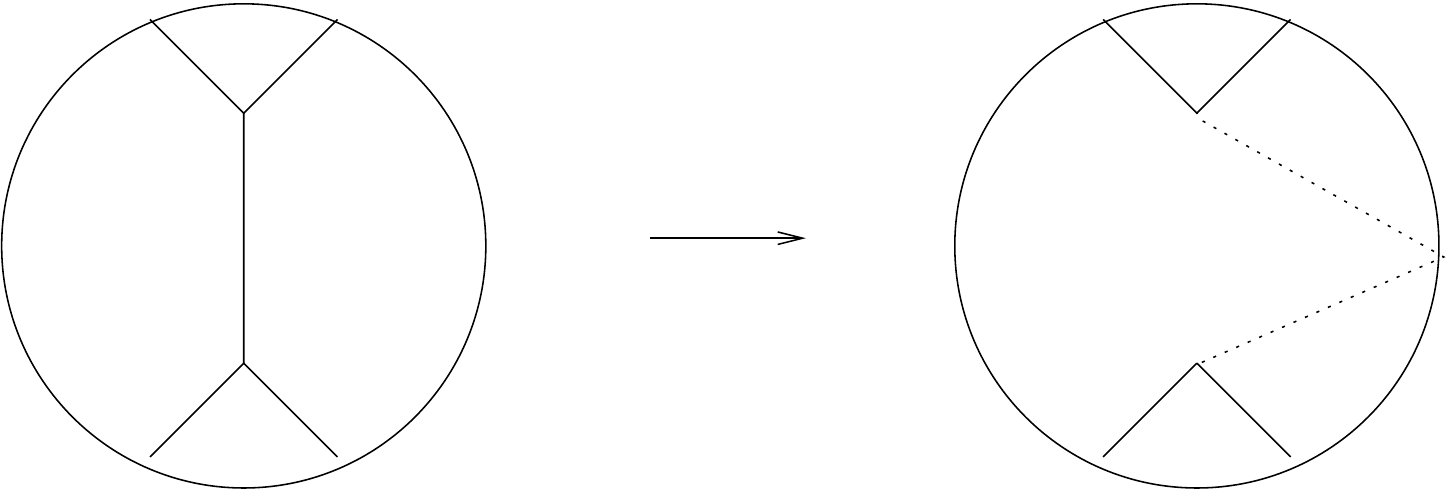}
\caption{\label{split} Recursion Relations}
\end{figure}

These recursion relations are shown schematically in Fig.\ref{split}.  Starting with a four-point vacuum correlator, we get the integral of the product of two three-point transition amplitudes each of which has one normalizable mode (shown by the dotted line).

It would be nice to find a way to relax our conditions on the polarization vectors. (In flat
space, with $\vect{\ep_1} = \vect{q}$, $\vect{\ep_n}$ can be arbitrary.) 
Nevertheless, by 
combining different BCFW extensions, we can compute
the amplitude
for arbitrary polarizations although we postpone the combinatoric details to a forthcoming paper.

\paragraph{\bf Gravity:}
Expanding gravity fluctuations about the background metric, ${\mathcal{G}}_{\mu \nu} = g_{\mu \nu} + h_{\mu \nu}$,
we find the gauge fixed quadratic Lagrangian \cite{Christensen:1979iy}
\begin{equation}
\begin{split}
S = &{-1 \over 64 \pi G} \int {d^d \vect{x} d z \over z^{d+1}} \left(\tilde{h}^{\mu \nu} \Box h_{\mu \nu} + 2 \tilde{h}^{\mu \nu} R_{\mu \rho \nu \sigma} h^{\rho \sigma} \right),
\end{split}
\end{equation}
where $\tilde{h}^{\mu \nu} = h^{\mu \nu} - {1 \over 2} g^{\mu \nu} h^{\alpha \beta} g_{\alpha \beta}$, 
and all covariant 
derivatives are with respect to the background metric.

It is now easy to extend the Ward identity argument above to gravity. We state the result. 
If we take the polarization vector for $\vect{k_1}$ to be
$\ep^1_{i j} = q_{i} q_{j}$, then the integrand of amplitudes dies off at large $w$ if we also take the polarization for $\vect{k_n}$ to be either (a) $\ep^n_{i j} = q_{(i}v_{j)},$ where $\vect{v} \cdot \vect{q} = 0$ or (b) $\ep^n_{i j}=v^1_{(i} v^2_{j)},$ where $\vect{v^m} \cdot \vect{q} = \vect{v^m} \cdot \vect{k_1} = 0$. 

With this constraint, the gravity recursion relations are the same as \eqref{gaugerecurs} with the obvious substitution of the  gauge polarization vectors, $\vect{\ep_m}$, with gravity polarization vectors. We will describe the
polarization-combinations that are accessible through combinations of BCFW extensions in a forthcoming paper. 

\paragraph{\bf Supersymmetric Theories:}
We can generalize these relations to supersymmetric theories as
in flat space \cite{Brandhuber:2008pf,*ArkaniHamed:2008gz, Lal:2009gn}.  \ads[d+1] supergroups do not 
exist for $d > 6$ and we are interested in the cases $d=4,5,6$ \cite{Nahm:1977tg}.  The reader may be more comfortable thinking about superconformal algebras in flat space although we will only use the super-Poincare subgroup of these algebras. A fact that we will use below is that \eqref{gaugerecurs} holds 
for Yang-Mills and gravity coupled to matter with the modification
that the sum over polarizations must be expanded to run over these particles
as well.

We explain the case for $d=4$ in detail and indicate results for $d=5,6$. In $d=4$, the 
superconformal group is $SU(2,2|{\cal N})$. For ${\cal N} = 4$, we have the 16 supercharges $Q^I_{\alpha},\bar{Q}_{\dot{\alpha} I}$ and their conformal 
partners. (We follow the conventions of  \cite{Kinney:2005ej}, so $I$ is an R-symmetry
index and $\alpha, \dot{\alpha}$ are spacetime spinor indices; see also \cite{Dolan:2002zh}.) We write the momenta we wish to extend, $\vect{k_1}$ and $\vect{k_n}$, as linear
combinations of two null vectors,$\vect{\lambda_1 \bar{\lambda}_1}$ and $\vect{\lambda_2 \bar{\lambda}_2}$, using  $(k_m)_{\alpha \dot{\alpha}} = \sum_{l=1}^2  a_{m l} \la_{l, \alpha} \lb_{l,\dot{\alpha}}$, (the $a_{m l}$ are some coefficients) and take $q_{\alpha \dot{\alpha}}= \la_{1 \alpha} \lb_{2 \dot{\alpha}}$. Next, we assemble the vector of $2 {\cal N}$-supercharges:  ${\cal Q}_{+}^A=\{\dotl[Q^I, \la_{2}], \dotlb[\bar{Q}_I, \lb_1] \}$. ($A$ runs over $1 \ldots 2 {\cal N}$.) For ${\cal N}=4$, defining $T_{--}=T_{i_1 j_1} q_{i_2} q_{j_2} \eta^{i_1 i_2} \eta^{j_1 j_2}$, 
we find that all states in the stress-tensor multiplet can be written
as (with $m=1$ or $m=n$)
\begin{equation}
\label{tcoherent}
T_m(\eta) = U_{+}(\eta) T_{--}(\vect{k_m}) U_{+}(-\eta); \, U_{+}(\eta) \equiv e^{{\cal Q}^A_{+} \eta_{A}}. 
\end{equation}

The expansion of these operators in the 8 Grassmann parameters $\eta_{A}$
contains all the original operators. With ${\cal N} = 2$, a similar expression exists for operators in
the same multiplet as a conserved current.


 We pause to note that only half-Bogomol'nyi-Prasad-Sommerfeld (BPS) multiplets of the superconformal algebra can be represented using a form like \eqref{tcoherent}. 
Although this form is not available for all half-BPS representations, 
it exists for all such representations
in $d=4,5,6$ that contain the stress tensor or a conserved current.

Now, consider a $n$-point correlator that involves two operators from \eqref{tcoherent} with
the {\em same} Grassmann parameter and $n-2$  other operators, which we denote below by the composite operator $O_{\mathcal C}$. The fact that this correlator is
invariant under supersymmetry transformations implies
\begin{equation}
\label{alltostress}
\langle T_1(\eta) T_n(\eta) O_{\mathcal C}  \rangle = \langle T_{--}(\vect{k_1}) T_{--}(\vect{k_n}) O^{\prime}_{\mathcal C} \rangle,
\end{equation}
where $O^{\prime}_{\mathcal C} \equiv U_{+}(-\eta) O_{\mathcal C} U_{+}(\eta)$. The right hand side 
can be computed by
BCFW recursion as explained above. 

So, 
supersymmetry allows us to compute a ``diagonal'' subset of correlators i.e correlators of operators in the stress-tensor multiplet where
at least two Grassmann parameters are the same. This suffices to 
determine the full set of 4-point correlators in ${\cn=4}$ SYM, 
 which can be reduced to one independent function \cite{Eden:2000bk,*Drummond:2006by}. But, in general, we would like to compute correlators where all Grassmann parameters are arbitrary.
This is possible with flat space amplitudes; the difficulty here 
is that we have stricter
constraints on the polarization-combinations that behave well 
under BCFW extension.

In $d=6$, the supercharges live in a 6 dimensional chiral-spinor representation (with eigenvalues ${\pm 1/2}$ under Lorentz transforms in 
the $(2i-1, 2i)$ plane) and 
in an R-symmetry group $Sp(2 {\cal N})$ where ${\cal N}$ is 1 or 2. (See \cite{Bhattacharya:2008zy} for conventions.)  
Apart from  the ``diagonal'' subset above, $d=6$ allows for
another calculable subset of correlators: 
We choose $\vect{k_1} = (1,0,0,0,0,0), \vect{k_n} = (a,b,0,0,0,0), \vect{q_1}=(0,0,0,0,1,I), \vect{q_n}=(0,0,1,I,0,0)$, and form two arrays of  $4 {\cal N}$  supercharges each: ${\cal Q}_{1 +}^A = \{Q^I_{ \pm 1/2, \pm 1/2, 1/2} \}$,  and ${\cal Q}_{n +}^A = \{Q^I_{\pm 1/2, 1/2, \pm 1/2} \}$. Then, for ${\cal N} = 2$,
with $T_{--}(\vect{k_m}) = T_{i j}(\vect{k_m}) q_m^i q_m^j$ and  $U_m(\eta)=\exp{[{\cal Q}^A_{m +} \eta_{A}]}$, we can compute any correlator of the form:
\begin{equation}
\label{sixcompute}
\left\langle U_1(\eta_1) U_n(\eta_n) T_{--}(\vect{k_1}) T_{--}(\vect{k_n}) O_{\mathcal C} U_n(-\eta_n) U_1(-\eta_1)\right\rangle.
\end{equation}
This 
is somewhat better than what we can do in $d=4$.

In $d=5$, the supercharges are spinors under $SO(5)$ and the R-symmetry $SU(2)$. This algebra has a half-BPS multiplet containing a conserved current and
we can compute diagonal correlators of operators in this multiplet.
However, the stress-tensor lives in a quarter-BPS multiplet \cite{D'Auria:2000ah}. 
So, not all operators in this multiplet can be reached via the analogue of \eqref{tcoherent} and we can only compute diagonal correlators among those
that can.  

\paragraph{\bf Results:}
We showed that transition amplitudes defined by \eqref{transdef}, which include vacuum correlators as a special case, could be calculated
by the recursion relations \eqref{gaugerecurs} for bulk Yang-Mills and gravity. Successive application of \eqref{gaugerecurs} allows us to relate all transition amplitudes to the three-point amplitude that is fixed, up to a constant factor, by conformal invariance.

These relations are also applicable to interacting bulk scalars with the
addition of a boundary term shown in \eqref{scalarecurs}. Supersymmetry allows us to compute additional correlators where
we can convert at least two operators to conserved currents or stress tensors
with appropriate polarizations. 
This includes the ``diagonal'' subset in \eqref{alltostress}
and, for $d=6$, also includes operators of the form \eqref{sixcompute}.

 In a forthcoming paper, we will apply these techniques to the calculation of higher order correlators. It would be very interesting to understand the deeper physical significance of these results and also extend them beyond tree-level in the bulk.

\paragraph{Acknowledgments:} I am grateful to S. Minwalla, R. Gopakumar, and  A. Sen for discussions. I acknowledge the support of a
Ramanujan fellowship and the Harvard University Physics Department.
\bibliography{references}
\end{document}